# Interface-enhanced ferromagnetism with long-distance effect in van der Waals semiconductor

*Wenxuan Zhu, Cheng Song,\* Lei Han, Hua Bai, Qian Wang, Siqi Yin, Lin Huang, Tongjin Chen, Feng Pan\**

W. Zhu, Prof. C. Song, L. Han, H. Bai, Q. Wang, S. Yin, Dr. L. Huang, T. Chen, Prof. F. Pan
Key Laboratory of Advanced Materials, School of Materials Science and Engineering, Beijing Innovation Center for Future Chips, Tsinghua University, Beijing 100084, China.
E-mail: songcheng@mail.tsinghua.edu.cn, panf@mail.tsinghua.edu.cn

Ferromagnetic semiconductors discovered in two-dimensional (2D) materials open an avenue for highly integrated and multifunctional spintronics. The Curie temperature ($T_C$) of existed 2D ferromagnetic semiconductors is extremely low and the modulation effect of their magnetism is limited compared with their 2D metallic counterparts. The interfacial effect was found to effectively manipulate the three-dimensional magnetism, providing a unique opportunity for tailoring the 2D magnetism. Here we demonstrate that the $T_C$ of 2D ferromagnetic semiconductor $Cr_2Ge_2Te_6$ can be enhanced by 130% (from ~65 K to above 150 K) when adjacent to a tungsten layer. The interfacial W-Te bonding contributes to the $T_C$ enhancement with a strong perpendicular magnetic anisotropy (PMA), guaranteeing an efficient magnetization switching by the spin-orbit torque with a low current density at 150 K. Distinct from the rapid attenuation in conventional magnets, the interfacial effect exhibits a weak dependence on $Cr_2Ge_2Te_6$ thickness and a long-distance effect (more than 10 nanometers) due to the weak interlayer coupling inherent to 2D magnets. Our work not only reveals a unique interfacial behavior in 2D materials, but also advances the process towards practical 2D spintronics.

## 1. Introduction

Ferromagnetic semiconductors, controllable by both magnetic and electric field, provide an efficient platform for the combination of spintronics and microelectronics.[1–4] Recently, ferromagnetic semiconductors were also discovered in the family of two-dimensional (2D) van der Waals (vdW) materials,[5,6] such as $Cr_2Ge_2Te_6$ (CGT),[7] $CrI_3$,[8] $CrBr_3$[9] and $MnBi_2Te_4$,[10] which exhibit abundant spintronic phenomena, including large tunneling magnetoresistance,[11–13] electric field tunable magnetism[14–17] and stacking-dependent interlayer coupling.[18–21] The flexible buildability of 2D ferromagnetic semiconductor also makes it a competitive candidate in integrated devices. Different from the enhanced ferromagnetism in their metallic counterparts,[22] the intrinsic Curie temperatures ($T_C$) of current vdW ferromagnetic semiconductors are below 100 K, which seriously impedes their practical applications. On account of the semiconductive property, electric field gating is the most commonly utilized to enhance the magnetism of 2D ferromagnetic semiconductors. However, the modulation efficiency of this method is quite limited in a very small scale, especially in CGT with $T_C$ around 65 K and weak PMA due to the Heisenberg behavior,[23] making the upper limit of the operation temperature in CGT-based devices only around 40 K.[24–26] Previous studies only showed the PMA modulation of CGT by ionic liquid gating with the extremely low $T_C$ unchanged[14]. In addition, by stronger electrostatic gating, the $T_C$ was raised but with extremely weak magnetic anisotropy (a kink in the magnetization curves) and the destruction of its semiconductive property.[16, 17] Therefore, effective methods for the strong modulation of vdW ferromagnetic semiconductors are still highly pursued.

The interfacial effects play a pivotal role in conventional magnetic films.[27,28] However, the interfacial effects in magnetic heterostructures are severely restrained by the thickness (*t*) and generally effective within a few nanometers at the interface following the 1/*t* principle.[29] For example, the proximity effect in (Ga,Mn)As/Fe bilayers can only extend to ~2 nm for the enhanced magnetic order in (Ga,Mn)As[30] and the electric field effect at the (La,Sr)MO$_3$/BaTiO$_3$ interface just works ~3 nm in the ferromagnetic (La,Sr)MO$_3$ layer.[31] The limited thickness for the interfacial effect seriously impedes the device design and applications. The interfacial phenomena in 2D magnets, which possess distinct interlayer (van der Waals) coupling from conventional magnetic heterostructures, would be fundamentally transformative.[32, 33]

In this work, through the construction of W/CGT heterostructure, we realized the elevation of the $T_C$ in CGT above 150 K with enhanced PMA by interfacial engineering. The interfacial W-Te bonding formed during the annealing treatment is responsible for the improvement of $T_C$ and PMA of CGT as supported by first-principle calculations. CGT remains semiconductive after $T_C$ enhancement and its magnetization can be effectively switched by spin-orbit torque (SOT) with a low current density ~2.4 × 10$^6$ A cm$^{-2}$ at 150 K. The weak interlayer coupling in 2D materials is also able to transfer the interfacial effect to much longer distance (>10 nm) compared to conventional three-dimensional (3D) systems in which the exchange coupling strength is isotropic. Besides the exclusive properties on nonmagnetic 2D (*e.g.* graphene[34], h-BN[35])/ferromagnet interfaces, the group of 2D "spinterface"[36] is extended to 2D ferromagnet/heavy metal system. Our work not only provides an effective method by interfacial engineering to manipulate the magnetism of 2D ferromagnetic semiconductor but also reveals the uniqueness of the interfacial effect in vdW materials.

## 2. Results and Discussion

The W/CGT bilayer was prepared by exfoliating ~10 nm CGT on 7 nm sputter-deposited W followed by annealing treatment at 400 °C for 4 hours. Then the sample was etched into a Hall bar with CGT at the center of the Hall cross for the measurements of out-of-plane magnetic field dependent Hall resistance ($R_H$–$H$). **Figure 1**a exhibits the schematics of the devices and setup for the Hall measurements, the magnetic field is applied along $z$-axis with the current along $x$-axis. Conventionally, due to the poor conductivity of CGT, the proximity to heavy metal is essential for the detection of magnetism of CGT by the anomalous Hall effect in the CGT/heavy metal bilayers.[37] The problem is that the formation conditions of proximity effect is strict, requiring etching process or exfoliation in vacuum for a fine interface.[24,37] The annealing process is also able to tightly bind the deposited W and exfoliated CGT and make the proximity-induced anomalous Hall effect detectable. Figure 1b shows the $R_H$–$H$ curves of annealed CGT at representative temperatures. Surprisingly, adjacent with W, the $R_H$–$H$ curves of CGT exhibit square shape with gradual shrink of coercivity and saturation magnetization from 30 K up to 150 K, which vanishes at 180 K, indicating a high $T_C$ above 150 K for CGT with strong PMA. For a comparison, Figure 1c displays the $R_H$–$H$ curves of ~10 nm-thick pristine CGT, which was measured by the proximity in CGT/Pt bilayer.[37] Slanted curves with abrupt attenuation up to 65 K (a kink) are observed in Figure 1c, indicating the weak PMA of CGT with $T_C$ around 65 K.[14, 23] Therefore, a combination of enhanced $T_C$ up to ~130% and PMA is achieved in W/CGT bilayer. A similar behavior is also observed in Ta/CGT bilayer (Figure S1, Supporting Information). In addition, the $T_C$-enhanced CGT remains semiconductive. As shown in Figure 1d, the vertical resistance of CGT after annealing with W, measured between the bottom W electrode and top Ti (10 nm)/Ag (70 nm) electrode increases with the decrease of temperature.

The persistence of semiconductive property in the $T_C$-enhanced sample reflects subtle doping effect.

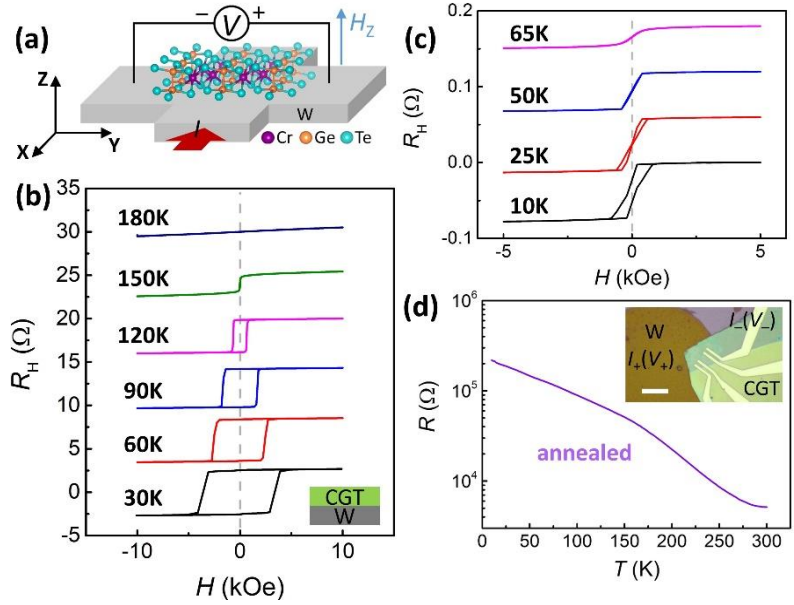

**Figure 1.** Transport measurements of $T_C$-enhanced CGT. a) Schematic view of the device for the measurement of $R_H$-$H$ curves. b,c) $R_H$-$H$ curves of b) W/CGT and c) pristine CGT at selected temperatures. The inset of b) shows the bilayer structure. d) $R$-$T$ curve of $T_C$-enhanced CGT. The inset shows the microscope image and measurement setup.

We then demonstrate that the enhanced ferromagnetism of CGT is the contributed by the interface of W/CGT bilayer. Firstly, the influence due to the change of crystal structure during the annealing process is ruled out by Raman measurements. The two characteristic peaks of the Raman spectra shown in **Figure 2**a represent two vibration modes of Cr-Te bonding in CGT and no shift is observed between pristine CGT and W/CGT heterostructure with CGT ~10 nm, demonstrating the unchanged crystal lattice of CGT. Then the interface of W/CGT bilayer was characterized through the high-angle annular dark-field scanning transmission electron microscopy (HAADF-STEM) imaging. Two areas with distinct contrast and an obvious interface are observed in the HAADF-STEM cross-sectional image shown in Figure 2b. The energy dispersive spectrometer (EDS) elemental mapping of the cross-section (Figure S2 Supporting Information) discloses that the top layered part is CGT and polycrystalline W occupies the bottom bright area. Due to the high atomization energy of tungsten, the diffusion of W atoms into CGT layers hardly happens at 400 °C, which is illustrated by the interface in EDS area mapping. Above the interface, the structure of one to two layers of CGT is distorted, shown by the obscure transition layer highlighted in Figure 2b, which is the results of the tight binding with heavy metal W at interface. The CGT layers above the interface is characterized to be uniform in both structure and composition (Section S2, Supporting Information), which is the same as the pristine CGT demonstrated by the Raman spectrum in Figure 2a. Therefore, the structural variation or Te-vacancy effect which can results in the regulation of magnetism[16,17,38] is excluded and we attribute the enhancement of PMA and $T_C$ to the interfacial effect. X-ray photoelectron spectroscopy (XPS) was then performed to directly demonstrate the interfacial W-Te bonding from the valence state. To rule out the influence of the bulk W and highlight the interface, W (2 nm)/CGT sample was prepared with identical annealing treatment and the XPS depth profile was obtained by Ar ion sputtering to compare the valence states of the interfacial W and the bulk W. As illustrated in Figure 2c, the sputtering with the time of 12 s and 48 s mainly collects the signal from the interface and the bulk W, respectively, according to the intensity of Te 4$d$ peaks. After 12 s sputtering, the coexistence of doublet peaks originated from W 4$f$ and Te 4$d$ indicates the signal of W 4$f$ is dominated by the interface. In contrast, with 48 s sputtering, the intensity of Te 4$d$ almost disappears with the simultaneous decline of the intensity of W 4$f$, suggesting the elimination of W/CGT interface and the signal of W 4$f$ mainly originates from the bulk. The most striking result is that the W 4$f$ peaks from the W/CGT interface shift by 0.375 eV to higher energy compared to the bulk, indicating the electron loss of W at the interface.[39,40] According to the electronegativity, W loses electrons and raises its valence when bonding with Te, which is in line with our XPS results. The thickness-dependent Raman spectrum is also investigated and a slight shift with widened peaks is only observed in much thinner samples, also indicating an interfacial effect rather than the variation of CGT structure, which can be ascribed to the formation of interfacial W-Te bonding (Figure S4, Supporting Information).[41]

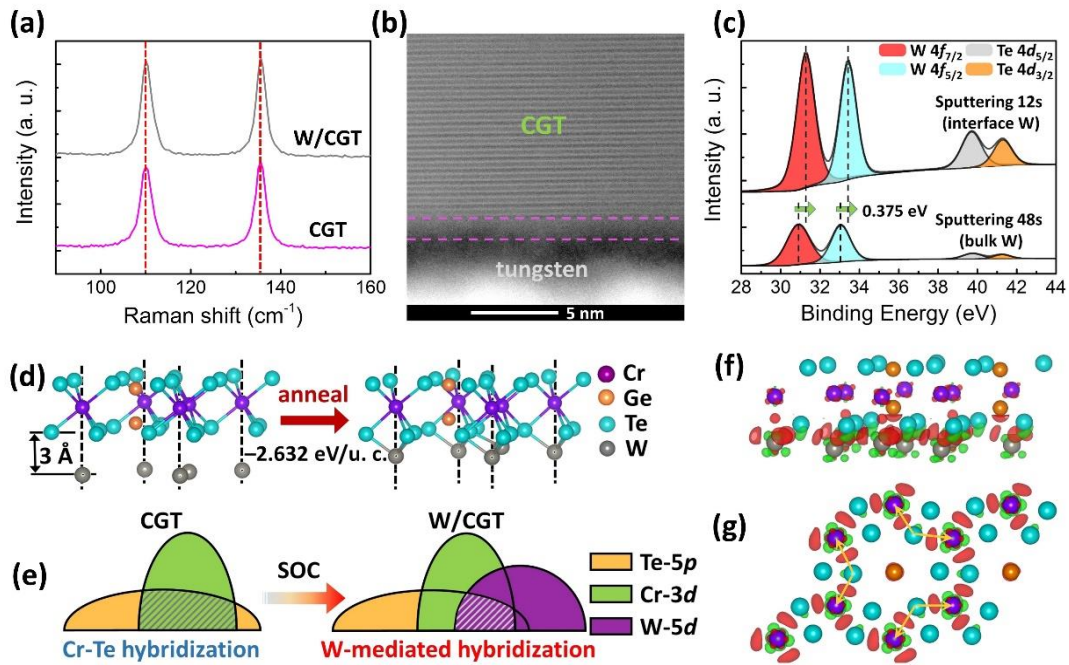

**Figure 2.** Characterizations and first-principle calculations of W/CGT bilayer. a) Raman spectra of pristine CGT and W/CGT with CGT ~10 nm. b) HAADF-STEM cross-sectional image of W/CGT bilayer. The transition layer between W and CGT is enclosed in purple dashed lines. c) Depth profile of W 4$f$ and Te 4$d$ high-resolution XPS spectra in W (2 nm)/CGT bilayer. d) Structural change of W/CGT heterostructure during the annealing process. The inset shows the energy change between initial and stable structure. e) Schematic of the orbital hybridization in pristine CGT and W/CGT heterostructure. f,g) DCD of W/CGT at interface from f)

the side view and g) top view (isosurface value of 0.0085 $e/$Bohr$^3$). The orange arrows represent the paths interacting two adjacent Cr atoms. Red and green isosurface contours indicate charge accumulation and reduction, respectively.

Then the mechanism of enhancement by the interfacial W-Te bonding is theoretically investigated through first-principle calculations. Figure 2d and 2e show the schematic of the physical picture. Firstly, the close adjacency compared to a gap (3 Å) between W and Te atoms at the interface will largely reduce the total energy by 2.632 eV/u. c. as shown in Figure 2d, which reflects the formation of W-Te bonding at the interface of deposited W and exfoliated CGT during the annealing process. The enhancement is attributed to the orbital hybridization, the schematic of which is shown in Figure 2e according to the calculations of density of states (DOS) (Figure S5, Supporting Information). In pristine CGT, the spin-orbit coupling (SOC) of Cr and the hybridization between Cr-3$d$ and Te-5$p$ orbitals dominants the strength of magnetic anisotropy and intralayer exchange coupling,[6] which determine $T_C$ of CGT. When W-Te bonding is formed in W/CGT heterostructure, the extended 5$d$ orbitals of W with strong SOC (about two orders of magnitude larger than CGT) participate in the hybridization, which enhances the SOC of adjacent CGT layer (Table S1, Supporting Information) with strengthened orbital interaction. By the calculations of magnetic anisotropy energy, it is found that the anisotropic exchange interactions induced by strong SOC of adjacent W together with the minor contribution of the enhanced ion anisotropy through increased SOC of Cr atom largely enhances the magnetic anisotropy (Table S2, Supporting Information). And the reconstructed orbital hybridization in CGT through the interfacial bonding between heavy metal W and Te, also results in the promotion of intralayer exchange coupling. The simultaneous enhancement of PMA and interlayer exchange coupling further lead to the increase of $T_C$.[42,43] The contribution of the CGT structural variation is also excluded. Detailed results of calculations are exhibited in Table S2. The doping effect induced by pure charge transfer is also a minor factor in W/CGT bilayer through the calculation of the Bader charge[44] and DOS (Section S4).

The interfacial W-Te bonding can be directly visualized by the differential charge density (DCD), which reveals the charge transfer at the interface. Figure 2f and 2g exhibits the side view and top view of the DCD. The strong interfacial W-Te bonding is reflected by the large charge redistribution at the interface, which significantly reconstructs the orbital hybridization interacting neighbored Cr atoms as illustrated by the large redistributed charge in the paths of superexchange coupling (Cr-Te-Cr) as highlighted in Figure 2g. The charge reduction around W shown in Figure 2f, demonstrates the electron loss of W when bonds with Te, consistent with the results of XPS. The reconstructed orbital hybridization also leads to the increase of unpaired electrons of Cr, as illustrated by the charge accumulation around Cr atoms in Figure 2e, which contributes to the increase of the magnetic moments (Table S2, Supporting Information).

The interfacial enhancement in 2D CGT also exhibits its uniqueness, which shows weak dependence on the thickness and can enhance the "bulk" magnetism. **Figure 3**a and 3b display the $R_H$–$H$ curves of W/CGT samples with two typical CGT thickness of 4.2 nm and 11.2 nm at selected temperatures from 90 K to 170 K, respectively. Note that although both thicknesses are enhanced by the interface, the 4.2 nm sample shows a slanted $R_H$-$H$ loop at 130 K with remanence ~50 % compared to the square loops in 11.2 nm sample, indicating the reduction of $T_C$ with weakened PMA. The variation of remanence with the thickness of CGT from 3.5 nm to 13.3 nm at 130 K is then summarized in Figure 3c. The remanence firstly increases with the thickness up to 7.7 nm and then persists at 100 % with further elevation of the thickness up to 13.3 nm showing a weak dependence on the thickness. The gradual weakening of magnetic anisotropy with the decrease of thickness below 7.7 nm, reflected by the reduced remanence, is originated from the enhanced thermal fluctuation and declined stability, which is quite characteristic for the few-layer vdW magnets. The squared magnetization hysteresis loop and obvious magnetic reversal of 13.3 nm sample in Figure 3d, obtained by magneto-optical Kerr effect (MOKE) at 150 K, further confirm that the interfacial enhancement is effective in relatively thick CGT. In contrast, the enhancement attenuates in the much thicker CGT (32.2 nm) with the absence of the hysteresis loop at 150 K, which is considered beyond the effective working distance of the interfacial effect. Note that the ferromagnetism in the samples with CGT below 4 layers is hard to be detected, which is due to the largely reduced stability, especially during the annealing and device fabrication process. It is then concluded that the interfacial enhancement in 2D CGT is effective beyond 10 nm. We attribute the anomalous interfacial effect to the weak interlayer coupling in 2D magnets. In 3D systems, due to the isotropic exchange coupling strength, the interfacial effect will be largely inhibited by the bulk effect and the propagation in the thickness direction is difficult.[28–30] The weak interlayer coupling, which is inherent to 2D materials, greatly debilitates the inhibition from the bulk effect and the interfacial effect can transfer a much longer distance.

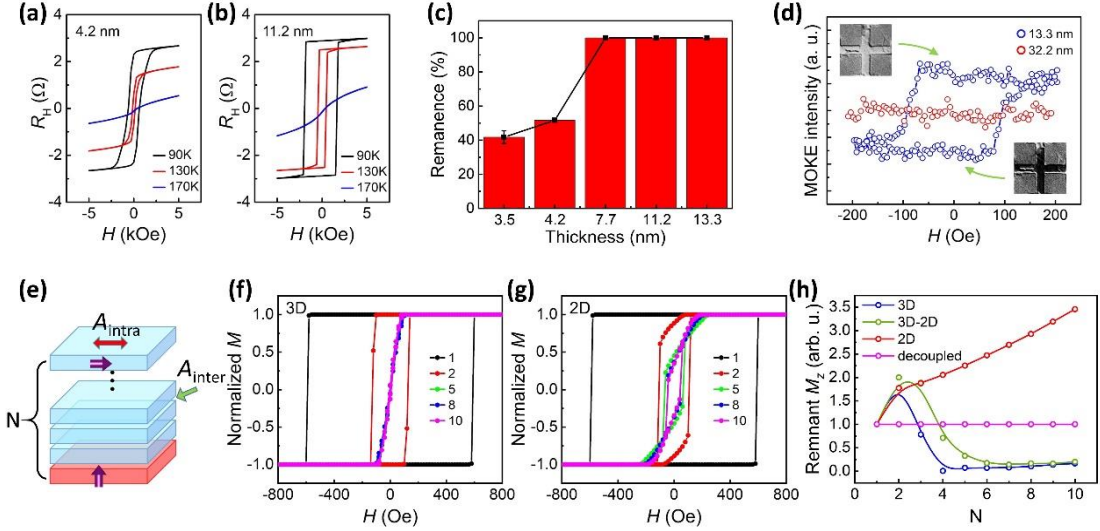

**Figure 3.** Transport, MOKE measurements and micromagnetic simulations of thickness-dependent magnetism. a,b) $R_H$-$H$ curves of W/CGT with a) 4.2 nm and b) 11.2 nm CGT at 90, 130 and 170 K. c) Variation of remanence with the CGT thickness measured at 130 K. d) Hysteresis loops of W/CGT with 13.3 nm and 32.2 nm CGT measured by MOKE at 150 K. The insets show the magnetic reversal between two states in 13.3 nm CGT. e) Schematic showing the layered model of interface enhanced PMA used for micromagnetic simulations. The purple arrows denote the magnetic moments. $A_{intra}$ and $A_{inter}$ represent the strength of intralayer and interlayer coupling, respectively. f,g) $M$-$H$ loops with different layer numbers N under f) $A_{inter} = A_{intra}$ (3D) and g) $A_{inter} = 0.04\ A_{intra}$ (2D). h) Variation of remanent $M_z$ with layers N in 3D, 3D-2D transition state, 2D systems and decoupled situation.

Then the micromagnetic simulations was utilized to reproduce the influence of interlayer coupling on the interfacial effect. A model of the interfacial enhanced PMA in multilayers imitating the vdW magnets is constructed. And the interlayer coupling strength is varied to investigate the change of the characteristic from conventional 3D to 2D systems. Figure 3e illustrates the schematic of the model. The PMA of the bottom layer is directly enhanced by the interface with $K_1$ = 0.2 MJ m$^{-3}$ and the magnetic anisotropy of the ground state in top layers is much smaller with $K_2$ = 0.01 MJ m$^{-3}$ referred to the first-principle calculations above. To enlarge the inhibition for the propagation of the interfacial effect and for better display of the effect of enhancement, the easy-axis of the top layers is set to be in-plane, rather than out-of-plane. $A_{intra}$ and $A_{inter}$ represent the strength of intralayer and interlayer coupling, respectively. Based on the first-principle calculation of vdW CGT, the strength of intralayer coupling is $A_{intra}$ = 2.368 pJ m$^{-1}$ with $A_{inter} = A_{intra}$ (3D) and $A_{inter}$ = 0.04 $A_{intra}$ (2D), representing the isotropic and anisotropic exchange coupling in 3D and 2D systems, respectively.

Figure 3f and 3g exhibit the variation of hysteresis loops with the number of layer N in 3D and 2D systems, respectively, and the behavior is highly related to the strength of the interlayer coupling. In conventional magnets (3D), the interfacial enhanced PMA is only effective below 5 layers and the magnetic anisotropy reverts to in plane in multiple layers indicating the invalid interfacial enhancement. By comparison, in 2D system, the clear hysteresis loop with a squared window is robust up to 10 layers, which exhibits barely no difference beyond 5 layers, indicating the effectiveness of interfacial induced PMA. The remanent out-of-plane magnetization ($M_z$) is then extracted from the simulated hysteresis loop to reflect the strength of the interfacial effect, which is summarized in Figure 3h. The rapid attenuation of the interfacial effect with layers N in 3D system is consistent with the common sense. In the transition state between 3D and 2D (3D–2D) with relatively smaller $A_{inter}$ (= 0.25 $A_{intra}$), the decline of $M_z$ is significantly slowed down with longer effective distance. Furthery, in 2D situation, a robust $M_z$ increases monotonically with N, indicating the long-term effectiveness of the interfacial induced PMA, which demonstrates the capability of the interlayer vdW coupling in the transfer of interfacial effect. Note that although the vdW coupling in 2D is much weaker than 3D situation, the interlayer coupling is still capable of interacting layers and dominating the interlayer magnetic coupling. Therefore, the 2D situation shows apparent difference with the decoupled situation ($A_{inter} = 0$), in which the $M_z$ is only contributed by the bottom layer and does not change with N. The remanence as percentage is shown in Figure S6 in Supporting Information.

Based on the interface enhanced-$T_C$ and PMA in CGT, we realize current-induced magnetization switching by SOT of CGT at 150 K. The measurement setup is shown in **Figure 4**a. An in-plane magnetic field aligned with the current is applied during the SOT switching to realize deterministic magnetization switching. We firstly show the $R_H$-$H$ curves with out-of-plane and nearly in-plane magnetic field measured at 150 K, as exhibited in Figure 4b. The much larger saturation field in the in-plane direction than the out-of-plane direction demonstrates a strong PMA at 150 K. The current-induced deterministic switching of 11.2 nm CGT is illustrated in Figure 4c. With 1 kOe in-plane external field applied, which is smaller than the anisotropic field shown in Figure 4b, the CGT is able to be efficiently switched by the current with two obvious states at zero current.[45] The polarity of the SOT switching is opposite when reverses the external magnetic field from 1 kOe to –1 kOe. The critical switching current density ~$2.4 \times 10^6$ A cm$^{-2}$ is obviously lower than other tungsten/ferromagnet bilayers[45–47] and is similar to topological insulators/CGT,[26] most likely ascribed to the formation of interfacial W-Te bonding. Note that the thermal effect generated by the current makes the temperature closer to $T_C$, resulting in the reduced Hall resistance of SOT switching to ~70 % of that in the $R_H$-$H$ curve.[48]

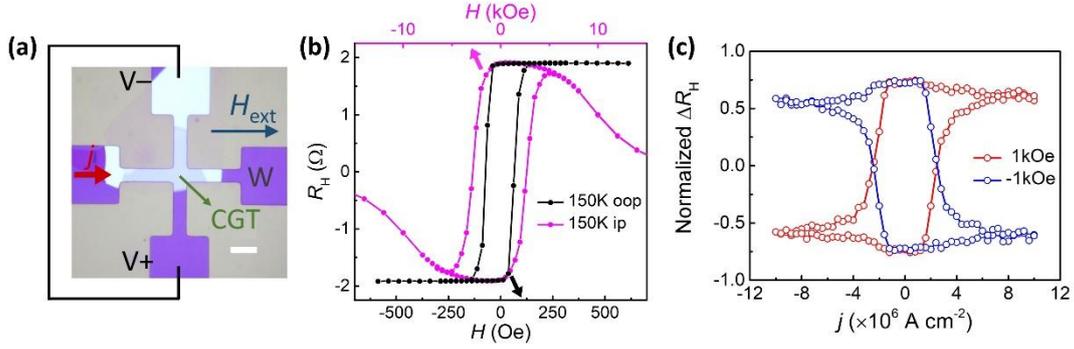

**Figure 4.** SOT of W/CGT bilayer at 150 K. a) Optical microscope image of the device and schematic of the SOT measurement. $H_{ext}$ and $j$ show the directions of applied external field and current. Scale bar: 5 μm. b) $R_H$-$H$ curves with applied out-of-plane (oop) and in-plane (ip) magnetic field at 150 K. c) Normalized $\Delta R_H$ as the function of current density applied to the device under 1 kOe and –1 kOe external field.

## 3. Conclusion

In conclusion, we reported that the $T_C$ and PMA of 2D semiconductor CGT can be remarkably enhanced by the interface with W, enabling the SOT switching of CGT with low current density at 150 K, which significantly promotes the practicability of 2D ferromagnetic semiconductors. The much longer effective distance of the interfacial enhancement in 2D CGT also reveals the capability of vdW gap in the transfer of interfacial effect, which is distinct from conventional magnetic films. The unique characteristic which can distinguish the vdW magnets from conventional magnets has been long-term pursued. The applicability of the interfacial engineering in other 2D magnetic semiconductors is also expected.

**Experimental Section**

*Sample Preparation*: W was deposited on Si/SiO2 substrate by magnetron sputtering with vacuum better than $10^{-7}$ Torr. CGT flakes were exfoliated using PDMS.

The heating process was also performed in magnetron sputtering system with vacuum better than $10^{-7}$ Torr. A Palladium layer ~3 nm was deposited above CGT for the protection from oxidation. For magneto-transport, 5 μm wide Hall bar devices were made by photolithography and ion milling.

*Characterizations of the structure and magnetism*: Raman analysis was carried out using a HORIBA Raman microscope with an excitation wavelength of 532 nm. The thicknesses of the samples were measured by atomic force microscope. Cross-section samples of the multilayer devices were fabricated by using a focused ion beam system. HAADF-STEM images, atomic resolved X-ray EDS and was performed on a FEI Titan Cubed Themis 18 60-300 (operated at 300 kV). Transport measurements were performed in Quantum design PPMS system. XPS The Kerr signal and images of magnetic reversal were captured by a Kerr microscope, which operates based on the magneto-optical Kerr effect (MOKE) in the polar configuration. The out-of-plane magnetization was probed and observed as different levels of brightness in the image. The measured binding energies of depth profile in XPS were corrected by referencing the Pd $3d_{3/2}$ peak for protection to 341.1 eV.

*First-principles calculations*: Our first principle calculations were performed using the Vienna *ab initio* simulation package (VASP)[49,50] with projector augmented wave method.[51,52] Perdew-Burke-Ernzerhof (PBE) functional[53] was used to treat the exchange correlation interaction and the plane-wave basis. The Gamma centered k-point mesh of $12 \times 12 \times 1$ was used in all calculations. A vacuum layer larger than 15 Å was adopted in all calculations of thin films. DFT-D3[54] was used to properly treat the interlayer van der Waals interaction. The GGA+U method was used to treat localized $3d$ orbitals, the $U_{\text{eff}}$ is selected to be 1.1 eV for the $3d$ orbitals of Cr according to previous study.[55] The unit cell of calculations is the primitive cell of CGT with 2 Cr atoms per unit cell.

*Micromagnetic simulations*: The micromagnetic simulations are conducted on the Object Oriented Micromagnetic Framework (OOMMF), which is a well-known open source software based on numerical solutions of the Laudau-Lifshitz-Gilbert equations. The discretized cell size equals to 0.68 × 0.68 × 0.68 nm, corresponding to the lattice constant of CGT. The saturated magnetization of both the interface enhanced layer and the other layers is set to be the same $M_s$ = 1.2 MA m$^{-1}$, with different PMA $K_1$ = 0.2 MJ m$^{-3}$ and in-plane $K_2$ = 0.01 MJ m$^{-3}$, respectively. Intralayer coupling $A_{intra}$ = 2.368 pJ m$^{-1}$ for all simulations and interlayer coupling $A_{inter}$ varies. The demagnetization energy is also included for accuracy. The initial magnetization state was set to be random, followed by relaxation and further stimulated by time-varying out of plane magnetic field $H$ for the simulation of magnetic hysteresis. A RungeKutta evolver was used to calculated the magnetization state with minimized energy under different $H$.

**Supporting Information**

Supporting information is available from the Wiley Online Library or from the author.

**Acknowledgments**

This work was supported by, the National Natural Science Foundation of China (Grant No. 51871130), and the Natural Science Foundation of Beijing, China (Grant No. JQ20010), and the National Key R&D Program of China (Grant No. 2017YFB0405704). C.S. acknowledges the support of Beijing Innovation Center for Future Chip (ICFC), Tsinghua University.

**Conflict of Interest**

The authors declare no competing interests.

**Author Contributions**

C. S. and F. P. led the project. W. X. Z. and C. S. proposed the study. W. X. Z. prepared the samples and carried out the measurements with the help from H. B. and S. Q. Y., W. X. Z. and L. H. conducted theoretical analysis. W. X. Z., L. H., H. B. and C. S. wrote the manuscript. All authors discussed the results and commented on the manuscript.

**Data Availability Statement**

The data that support the findings of this study are available from the corresponding author upon reasonable request.